\documentstyle[preprint,prc,aps]{revtex}
\begin{document}
\preprint{
\rightline{\vbox{\hbox{\rightline{MSUCL-1028}} 
\hbox{\rightline{nucl-th/9604034}}}}
         }
\title{Coherent Pion Pairs from Heavy-Ion Collisions}
\author{Scott Pratt and Kevin Haglin}
\address{Department of Physics and Astronomy and National Superconducting
Cyclotron Laboratory, Michigan State University, East Lansing, MI 48824-1321,
USA}
\date{\today}
\maketitle
\begin{abstract}
The degree to which a nucleus can act as a source for coherent pion pairs is
investigated for intermediate-energy heavy-ion collisions.  Creation through
both isovector and isoscalar channels is considered.  Two experimental signals
are proposed for evidence of two-pion coherent production, two-pion enhancement
and the focusing of outgoing pions along the beam axis.
\end{abstract}

From a variety of perspectives, coherent pion emission has attracted a great
deal of attention.  Purely theoretical issues regarding coherent
sources\cite{andreev,rudypaper} have been investigated, as well as in the
context of disoriented chiral condensates, CENTAURO cosmic ray
phenomenology\cite{wilczek,gavin}, and subthreshold pion emission of single
pions\cite{vasak,prakash,benenson,stachel,miller}. The difficulty in physically
justifying coherent pion sources stems from the facts that pions are
pseudoscalars and that they carry isospin.  Thus a classical source needs to
have the same characteristics.  It is acceptable to treat quantities such as
the electric field of an ion or the nuclear mean field as classical entities.
However, operators which carry conserved quantities like charge or angular
momentum should either be treated as transition operators, or should be
considered only in the context of higher order fluctuations\cite{vasak} and
correlations\cite{gavin}.  Such considerations are difficult to consistently
incorporate in theoretical models. The goal of this paper is to explore the
coherent emission of pion pairs, which can indeed have a coupling to either the
net baryon density (through the $\sigma$ channel), or to the neutron excess of
a nucleus (through the $\rho$ channel).

For either channel we will work with an effective four-point coupling of the
pions to the nucleons, which will be translated as a coupling to the
appropriate density $j(x)$ in the following form.

\begin{equation}
{\cal L}_{\rm \, eff} = j(x) {\cal M}(k_1,k_2,n) {\vec \pi}\cdot {\vec \pi},
\end{equation}
where $n^{\mu}$ is a unit four vector representing the rest frame of the
matter, $j(x)$ represents the density in the frame of the matter, and $k_1$ and
$k_2$ denote the momenta of the outgoing pions.

Given ${\cal L}_{\rm \, eff}$, one can find the two-pion emission probability:
\begin{equation}
\frac{dN}{d^3k_1d^3k_2}=\frac{1}{(2\pi)^6 2E_12E_2}
\left | \int d^4x {\cal M}(k_1,k_2,n) j(x) e^{i(k_1+k_2)\cdot x}
\right |^2
\end{equation}

First, we consider the isoscalar channel.  At low momentum, isoscalar pion
pairs do not couple to nuclear matter due to chiral
symmetry\cite{kapusta,matsui}, which is demonstrated by the extremely small
pion-nucleon scattering length.  However, p-wave couplings of pions to nucleons
is strong\cite{bhaduri}:
\begin{eqnarray}
{\cal L_{\pi N}} &=& \frac{f}{m_{\pi}} \bar{\psi}_N\gamma^{\mu}
\gamma^5({\vec \tau})\psi_N\cdot\partial_{\mu}{\vec \pi},\\
\nonumber
\frac{f^2}{4\pi} &=& 0.08
\end{eqnarray}
Drawing a graph with two such vertices (See Fig. \ref{feynmannfig}), and making
the approximation that the mass of the nucleon is large compared to the pion
mass and that the motion of the baryons in nonrelativistic, one derives an
effective four-point coupling for the coupling of an isoscalar pair of pions
coupled to baryons.
\begin{equation}
{\cal M}(k_1,k_2,n) = \frac{f^2}{m_{\pi}^2} \left\{\frac{1}{n\cdot k_1}
+\frac{1}{n\cdot k_2}\right\}
\left\{k_1\cdot k_2 -\frac{(k_1\cdot n)(k_2\cdot n)}{n^2}\right\}.
\end{equation}
The appropriate density $j(x)$ for the scalar case is the baryon density.

For emitting through the isovector channel, we consider the simple process of
two pions coupled to a nucleon through the $\rho$ meson (See Fig. \ref{feynmannfig}), which leads to the
expression:
\begin{equation}
{\cal M}(k_1,k_2,n) = g_{\rho}^2 
\frac{(k_1-k_2)\cdot n}{(k_1+k_2)^2-m_{\rho}^2+im_{\rho}\Gamma }
\end{equation}
The current $j(x)$ one would couple to here would be one half the difference of
the neutron and proton densities.

%\begin{figure}
%\begin{center}
%\epsfig{file=figs/feynmann.eps,width=6cm}
%\end{center}
%\caption[]{The upper figure represents the coupling of two pions to a baryon
%through two derivative couplings, resulting in an effective isoscalar coupling.
%The lower figure illustrates two pions coupling to a nucleon through the $\rho$
%meson.}
%\label{feynmannfig}
%\end{figure}

As long as the nucleons remain non-relativistic, $n$ can be considered
independent of $x$ and ${\cal M}$ can be factored out of the integral over $x$.
For our calculations, $n$ is considered to be time-like in the center-of-mass
frame of the colliding nuclei. For one nucleon which moves along a
straight-line trajectory from $x_1$ to $x_2$, the integral over $x$ becomes:
\begin{eqnarray}
\int d^4x e^{iK\cdot x} j(x) &=& \frac {i}{K_0 - {\bf K}\cdot {\bf v}}
\left(e^{iK\cdot x_1}-e^{iK\cdot x_2}\right),\\
\nonumber
{\bf v} &=& \frac {{\bf x}_2-{\bf x}_1}{t_2-t_1}.
\end{eqnarray}
From one element of a trajectory to the next, the contributions cancel as long
as the velocity does not change.

If one knows the behavior of the baryon density as a function of $x$, and if
one can approximate such behavior with continuous trajectories of
representative baryonic charges, one can predict the coherent component of the
emission by summing over the trajectories.  Although the expression above is
rather straight-forward and the baryon density as a function of time can be
obtained easily from simulations, implementing the expression is difficult.  In
principle, random back and forth Fermi motion cancels out in the integral, but
the cancelation is only statistical in a simulation and the noise due to the
nucleons scattering and reflecting off the mean field makes a straight-forward
implementation of the trajectories untenable.

Any set of trajectories that describes the evolution of the baryon density as a
function of time will be suitable.  To minimize the random motion of the
trajectories, we consider the trajectory of a point, $z(t), r_{\perp}(t),
\phi(t)$, which stays ahead of a fraction, $f_z$ of the baryons in the
$z$-direction, while staying inside a fraction $f_{\perp}$ of baryons at that
$z$ position.  By sampling trajectories with random values of $f_z$ and
$f_{\perp}$ between zero and one, and with random values of the azimuthal angle
$\phi$, the evolution of $\rho({\bf r},t)$ is represented. For a fixed
$f_{\perp}$ and $f_z$, a trajectory is defined as:
\begin{eqnarray}
2\pi\int_{z^{\prime}<z} dz^{\prime} \rho(r_{\perp}^{\prime},z^{\prime}) r_{\perp}^{\prime} dr_{\perp}^{\prime}
&=& f_z\cdot A\\
\nonumber
\int_{r_{\perp}^{\prime}<r_{\perp}} \rho(r_{\perp}^{\prime},z) r_{\perp}^{\prime} dr_{\perp}^{\prime}
&=&f_{\perp} \int \rho(r_{\perp}^{\prime},z) r_{\perp}^{\prime}
dr_{\perp}^{\prime}
\end{eqnarray}
To calculate such trajectories, particle densities were calculated as a
function of time from a BUU simulation of a zero-impact-parameter 150$\cdot$A
MeV collision of 100 nucleons on 100 nucleons using 2000 test particles per
nucleon.  Each 2 fm/c, boundaries which were perpendicular to the beam axis
were found such that 5\% of the particles where in each region.  Within each
slice, five regions were defined transversely as rings with one fifth of the
particles in each ring.  Trajectories were calculated for points which stayed
in the same region, e.g. the $n^{th}$ slice along the beam and in the $i^{th}$
ring of that slice, and which stayed at a fixed percentage between the
boundaries and at a fixed azimuthal angle.  The trajectories are shown in Fig. \ref{trajectoryfig}, while a sampling of phase space points for the trajectories at various times are illustrated in Fig. \ref{evolutionfig}.

%\begin{figure}
%\begin{center}
%\epsfig{file=figs/trajectory.eps,width=7.5cm}
%\end{center}
%\caption[]{The trajectories of various points of given values of $f_z$ and five
%different values of $f_{\perp}$ are illustrated in each panel.  Positions are
%shown in center of mass coordinates for a symmetric central collision of
%$A=100$ nuclei at $E/A =$ 150 MeV.}
%\label{trajectoryfig}
%\end{figure}

%\begin{figure}
%\begin{center}
%\epsfig{file=figs/evolution.eps,width=7.5cm}
%\end{center}
%\caption[]{A sampling of phase points used to calculate the trajectories in
%Fig. \ref{trajectoryfig} are shown for several times.}
%\label{evolutionfig}
%\end{figure}

The fourier transform of the trajectory was calculated as shown above, and pion
spectra were obtained.  Several million trajectories were numerically sampled,
and the matrix elements were added coherently.  For each point, an absorption
coefficient for the matter was estimated which depended on the probability of a
particle escaping from a sphere of nuclear matter, with the point from the
sphere at the same fractional position both along the beam and radially.  Thus,
the absorption was independent of time, significantly simplifying the
calculation.  A three-fermi absorption distance was assumed.

Unfortunately, the magnitude of the signal was rather sensitive to the time
step chosen.  Calculations were performed with .5 fm/c time steps, with the
trajectories being splined from their 2 fm/c values. This splining somewhat
reduced the sharpness of the deceleration.  Of course, one could extract points
from BUU at smaller time intervals, but such small steps might yield boundaries
with rapid statistical fluctuations, which might unphysically radiate.  By
studying the behavior of the emission on the number of test particles, it was
determined that radiation was not due to numerical fluctuations. At 150$\cdot
A$ MeV, for the isoscalar case, the number of charged pions per event came out
as 1 per $10^5$, while thirty percent as many were created at 100$\cdot A$ MeV.
For the isovector case, at 150$\cdot A$ MeV and assuming 42 protons and 58
neutrons in each nucleus, there were only one third the number of charged
pions and no neutral pions.

Although we still suspect that the magnitude of the signal might change by as
much as a factor of two due to details of the calculation, one signature of the
spectra was completely robust.  Pion momenta were focused along the beam axis.
Fig. \ref{angdistfig} demonstrates this by showing the angular distribution of
the the pion's momentum for pions with energies below 100 MeV, and pions with
energies above 100 MeV.  The angular distribution of the sum of the two momenta
is also shown for both the isoscalar and isovector cases. For the isovector
case the pair momenta was not so focused due to the $\rho$ form factor which
preferred pion pairs with a large invariant mass. Calculations never included
pions with kinetic energies above 200 MeV. A form factor would have more
realistically incorporated the cutoff, but would have slowed the Monte Carlo
procedure.  The single-pion energy spectra peaked near 100 MeV.

%\begin{figure}
%\begin{center}
%\epsfig{file=figs/angdist.eps,width=8cm}
%\end{center}
%\caption[]{The angular distribution of pions and pion pairs is shown as a
%function of the center-of-mass angle for $150\cdot A$ MeV symmetric collisions
%of A=100 nuclei.  Distributions for pions with kinetic energies below 100 MeV
%(squares), above 100 MeV (triangles) and distributions for the sum of the pion
%momenta (circles) are shown.  The left/right panels are for the
%isoscalar/isovector case respectively.}
%\label{angdistfig}
%\end{figure}

The focusing can be understood by considering the phase factor $e^{iK\cdot x}$
in the expression above.  Accelerations only emit coherently when relative
phases are small compared to $\pi$.  Since $K_0$ is near three hundred MeV,
only accelerations that occur within one or two fm/c of one another will
interfere constructively.  The strongest deceleration occurs from the shock
wave that travels along the beam axis from the point of initial impact.  Points
that emit coherently are confined to a slice perpendicular to the z-axis by the
constraint that accelerations happen at the same time.  For the same reason
that light from a circular aperture will be focused forward, pion pairs are
then focused along the z-axis due to the planar shape of the shock wave at a
given time.

These calculations could be improved in a variety of ways.  Other sources of
emission need to be considered.  Emission through the $\rho$ meson can be
responsible for pion pairs but only for heavy nuclei where there is a
neutron/proton imbalance.  The treatment of absorption presented here is surely
oversimplified.  Correlations in the nucleus might increase the strength of
coherent emission.  A more correct quantum treatment would replace the
description of the emission as coming from particular space-time points, with a
description of emission from particular nuclear states. It is an especially
daunting challenge to accomplish these goals for a non-zero impact parameter
example.

Despite the theoretical difficulty in calculating the yield, the interpretation
of experimental results has the attraction of being definitive in two aspects.
First, if focusing is observed in the pion spectra along the beam axis it
signals coherent emission.  In central collisions of large symmetric nuclei,
single-particle emission would peak transverse to the beam due to shadowing.
The effects of shadowing had been expected from model calculations
{\cite{bauer,qmd} and have been experimentally verified \cite{gsi}.  Since the
coherent component is expected to be at higher average energy than the usual
subthreshold contribution, the focusing should be more obvious when gating on
higher-energy pions.  In fact, backwards focusing has been
observed\cite{schubert} in Ar+Au reactions, and the focusing has been seen to
be stronger for more energetic pions.  However, to be convinced that it is not
shadowing, the experiment needs to be done with a symmetric system, and with a
tight centrality cut.

A second and more obvious signature is to search for two-pion coincidences. An
enhancement of the two-pion yield relative to the one-pion yield squared would
signal isoscalar emission if it were in both the neutral and charged-pion
measurements, and would signal coherent emission through the $\rho$ channel if
it showed up in only the charged pion measurements.  The experimental
challenges are significant as one the measurement of $10^9$ central events is
required along with good phase space coverage for pions.  It would be
especially difficult to measure both positive and negative pions in the same
experiment.

%%%%%%%%%%%%%%%%%%%%%%%%%%%%%%%%%%%%%%%%%%%%%

\acknowledgments{The authors wish to thank Frank Daffin and Wolfgang Bauer for supplying the BUU code used in this work.  This work was supported by the National Science Foundation through grant no. PHY-9403666.}

%\end{document}

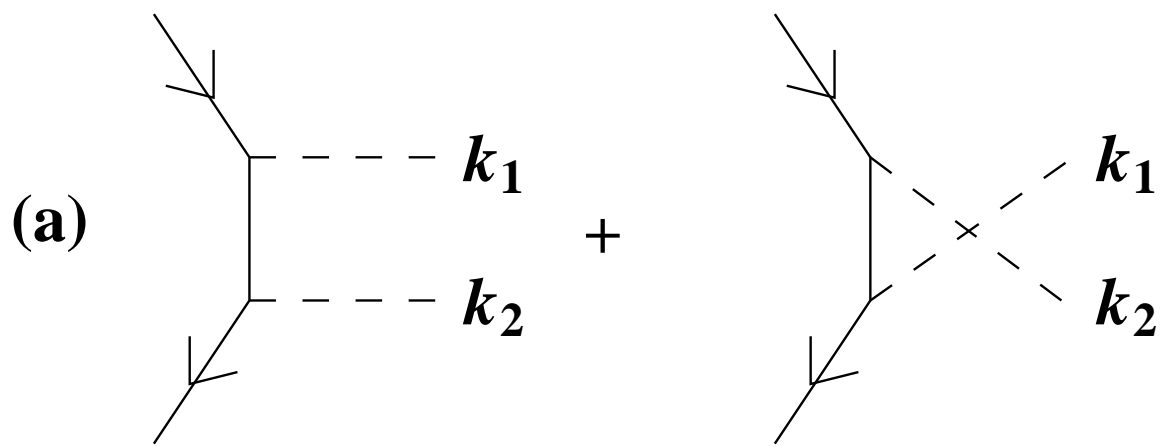
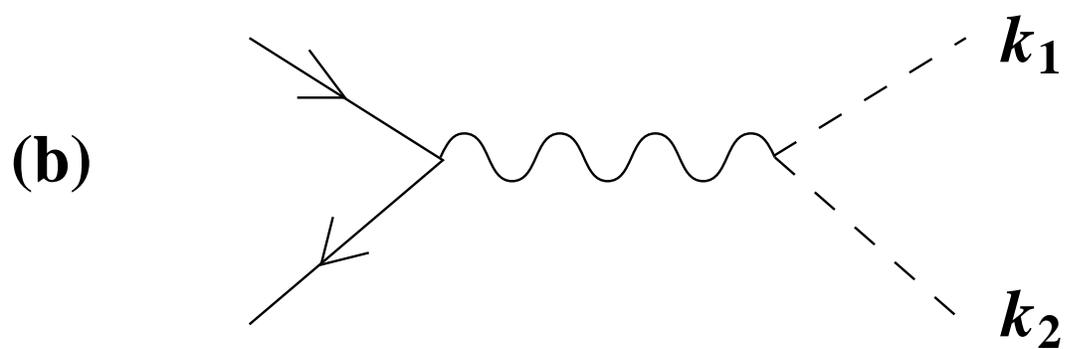
\begin{figure}
\caption[]{The upper figure represents the coupling of two pions to a baryon
through two derivative couplings, resulting in an effective isoscalar coupling.
The lower figure illustrates two pions coupling to a nucleon through the $\rho$
meson.}
\label{feynmannfig}
\end{figure}

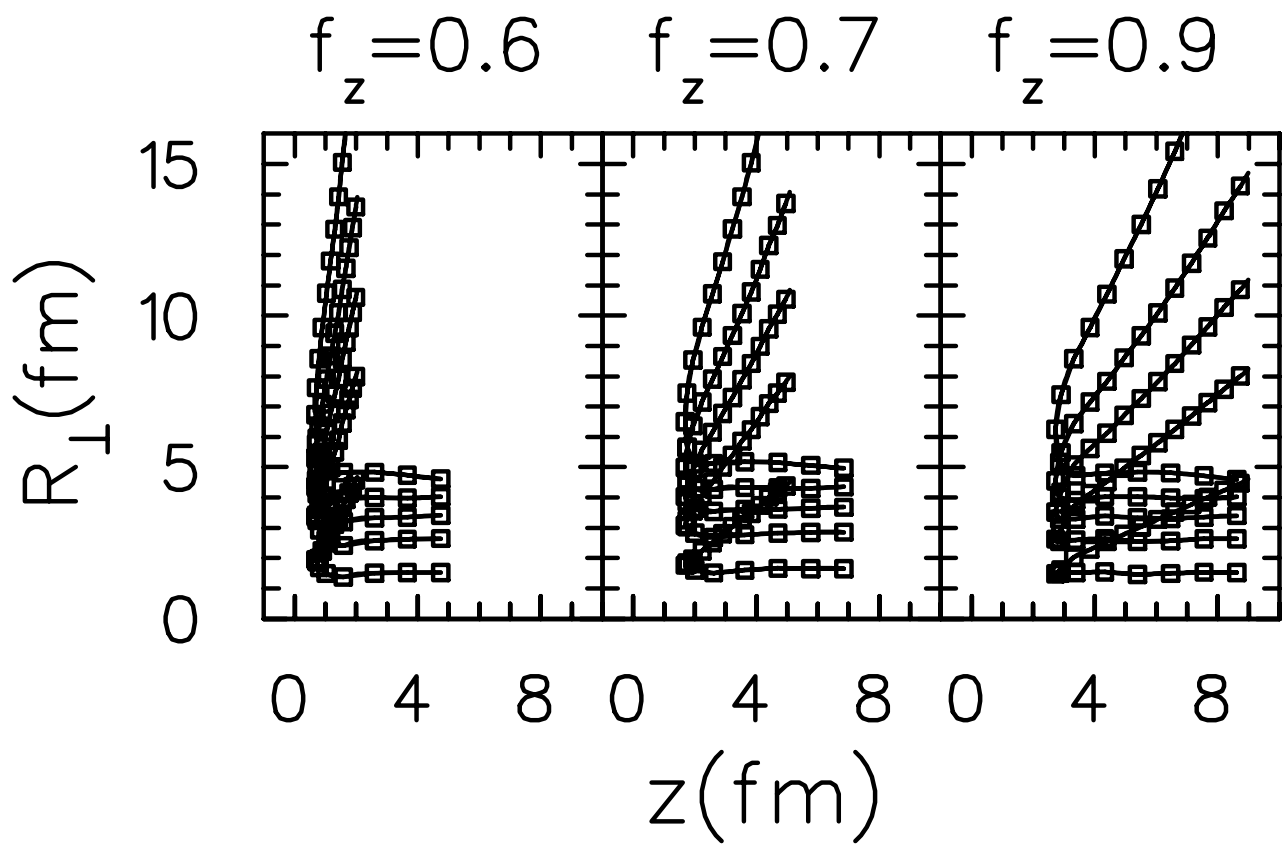
\begin{figure}
\caption[]{The trajectories of various points of given values of $f_z$ and five
different values of $f_{\perp}$ are illustrated in each panel.  Positions are
shown in center of mass coordinates for a symmetric central collision of
$A=100$ nuclei at $E/A =$ 150 MeV.}
\label{trajectoryfig}
\end{figure}

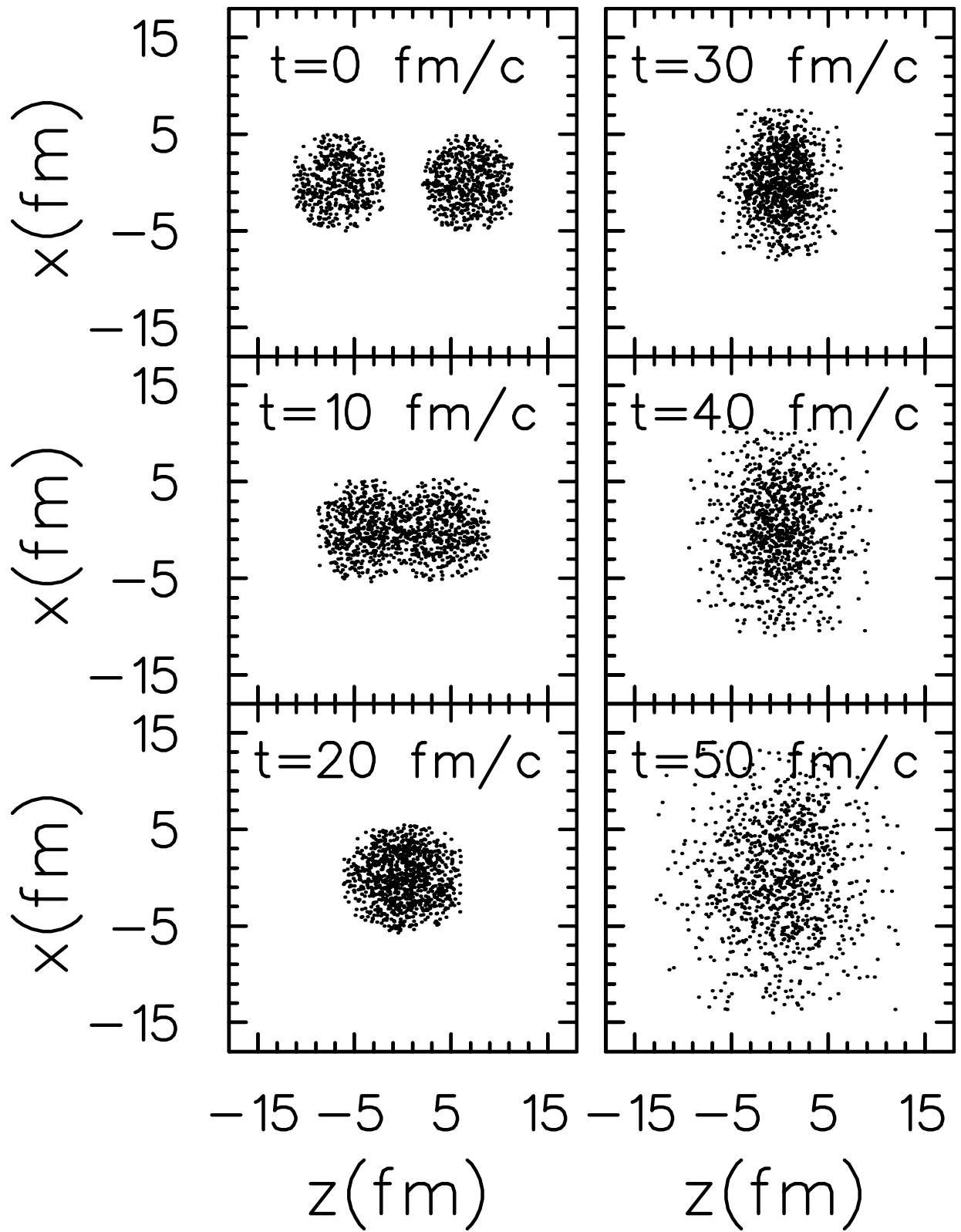
\begin{figure}
\caption[]{A sampling of phase points used to calculate the trajectories in
Fig. \ref{trajectoryfig} are shown for several times.}
\label{evolutionfig}
\end{figure}

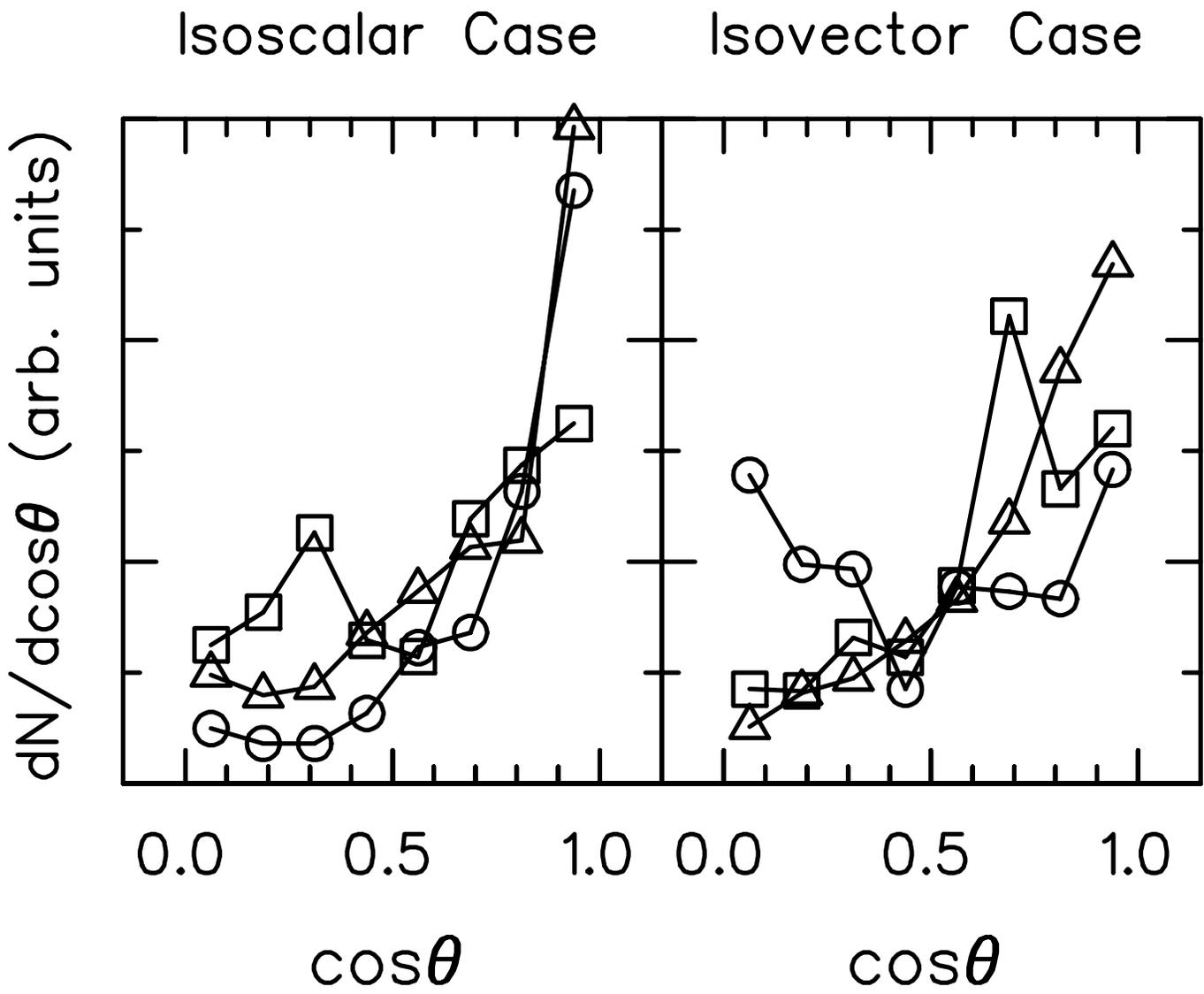
\begin{figure}
\caption[]{The angular distribution of pions and pion pairs is shown as a
function of the center-of-mass angle for $150\cdot A$ MeV symmetric collisions
of A=100 nuclei.  Distributions for pions with kinetic energies below 100 MeV
(squares), above 100 MeV (triangles) and distributions for the sum of the pion
momenta (circles) are shown.  The left/right panels are for the
isoscalar/isovector case respectively.}
\label{angdistfig}
\end{figure}


\begin{references} 

%\bibitem{mcmillan} McMillan and E. Teller, Phys. Rev. {\bf 72}, 1 (1947).

%\bibitem{heisenberg}, W. Heisenberg, Z. Phys.  {\bf 126}, 569 (1949).

\bibitem{andreev} I.V. Andreev, M. Pl\"umer and R.M. Weiner, Phys. Rev. Lett.
{\bf 67}, 3475 (1991).

\bibitem{rudypaper} S. Pratt, ``Quark-Gluon Plasma 2'', ed. R. Hwa, World Scientific (1996).

\bibitem{wilczek} K. Rajagopal and F. Wilczek, Nucl. Phys {\bf B399}, 395
(1993).

\bibitem{gavin} S. Gavin and B. M\"uller, ``Larger Domains of Disoriented
Chiral Condensate Through Annealing,'' Phys.\ Lett.\ {\bf B329}, 486 (1994).\\
S. Gavin, A. Gocksch and R.D. Pisarski, ``QCD and the Chiral Critical Point,''
Phys.\ Rev.\ {\bf D49} R3079 (1994).\\
S. Gavin, A. Gocksch and R.D. Pisarski, Phys.\ Rev.\ Lett.\ {\bf 72}, 2143 (1994).

\bibitem{vasak} D. Vasak, W. Greiner and B. M\"uller, Nucl. Phys. {\bf A428}, 291c (1984).

\bibitem{prakash} M Prakash, C. Guet and G.E. Brown, Nucl. Phys. {\bf A447} 625c (1985).

\bibitem{benenson} W. Benenson, {\sl et al.}, Phys. Rev. Lett. {\bf 43}, 683 (1979).

\bibitem{stachel} J. Stachel, {\sl et al.}, Phys. Rev. C{\bf 33}, 1420 (1986).

\bibitem{miller} J. Miller {\sl et al.}, Phys. Lett. {\bf B314}, 7 (1993).

\bibitem{kapusta} J.I. Kapusta, \underline{Finite-Temperature Field
Theory},Cambridge University Press, p. 193, (1989).

\bibitem{matsui} T. Matsui and B.D. Serot, Ann. Phys. {\bf 144}, 107 (1982).

\bibitem{bhaduri} R.K. Bhaduri, \underline{Models of the Nucleon},
Addison-Wesley, p. 134, (1987).

\bibitem{bauer} B.-A. Li, W. Bauer and G.F. Bertsch, Phys. Rev. C{\bf 44},
2095, (1991).

\bibitem{qmd} Ch. Hartnack, {\sl et al.}, Nucl. Phys. {\bf A538}, 53c (1992).

\bibitem{gsi} L.B. Venema, {\sl et al.}, Phys. Rev. Lett. {\bf 71}, 835 (1993).

\bibitem{schubert} A. Schubert {\sl et al.}, Phys. Lett. {\bf B328}, 10 (1994).

\end{references}
\end{document}